\documentclass[aps,prb,twocolumn]{revtex4}
\usepackage{amssymb}
\usepackage{amsmath}
\usepackage{graphicx}

\begin{document}

\title{Sign reversal superconducting gaps revealed by phase referenced quasi-particle interference of impurity induced bound states in (Li$_{1-x}$Fe$_x$)OHFe$_{1-y}$Zn$_y$Se}

\author{Qiangqiang Gu, Siyuan Wan, Zengyi Du, Xiong Yang, Huan Yang$^{*}$, Hai Lin, Xiyu Zhu, and Hai-Hu Wen$^{\dag}$}

\affiliation{National Laboratory of Solid State Microstructures and Department of Physics, Collaborative Innovation Center of Advanced Microstructures, Nanjing University, Nanjing 210093, China}

\begin{abstract}
By measuring the spatial distribution of differential conductance near impurities on Fe sites, we have obtained the quasi-particle interference (QPI) patterns in the (Li$_{1-x}$Fe$_x$)OHFe$_{1-y}$Zn$_y$Se superconductor with only electron Fermi surfaces. By taking the Fourier transform on these patterns, we investigate the scattering features between the two circles of electron pockets formed by folding or hybridization. We treat the data by using the recent theoretical approach [arXiv:1710.09089] which is specially designed for the impurity induced bound states. It is found that the superconducting gap sign is reversed on the two electron pockets, which can be directly visualized by the phase-referenced QPI technique, indicating that the Cooper pairing is induced by the repulsive interaction. Our results further show that this method is also applicable for data measured for multiple impurities, which provides an easy and feasible way for detecting the gap function of unconventional superconductors.
\end{abstract}

\maketitle

\section{INTRODUCTION}

The discovery of iron based superconductors\cite{HosonoJACS2008} provides us a second example of unconventional high temperature superconductors. It is categorized as ``unconventional'' because a lot of unique features have been found. For example, the parent phase of $R$FeAsO  ($R$ = rare earth elements) and $AE$Fe$_2$As$_2$ ($AE$ represents the alkaline earth metals Ba, Sr, Ca, etc.) have the long range antiferromagnetic (AFM) orders \cite{DaiPCNature2008,JohrendtPRL2008}. The superconductivity emerges when this long range AFM order is suppressed. Plenty of evidence indicates that superconductivity has been mediated by AFM spin fluctuations in the pairing process\cite{ImaiPRL2014}. Theoretically it was proposed that the pairing may be established by the pair-scattering of two electrons with opposite momenta between the hole and electron pockets\cite{Mazin,Kuroki} leading to the so-called sign reversal $s$-wave gap, namely the $s^\pm$-pairing manner. This picture, originally proposed for the FeAs-based system with both electron and hole pockets, has been actually supported by several important experiments, such as the quasiparticle interference (QPI) in FeTe$_{1-x}$Se$_x$\cite{Hanaguri} and the inelastic neutron scattering\cite{NatureNeutronScattering}. We have also done the experiments of scanning tunneling microscopy/spectroscopy (STM/STS) measurements on the non-magnetic Cu impurities in Na(Fe$_{0.96}$Co$_{0.03}$Cu$_{0.01}$)As and found clear evidence of the in-gap bound states providing strong support of the $s^\pm$-pairing\cite{YangHNC2013}. In addition, a bosonic mode was observed outside the superconducting coherence peaks in at least two systems\cite{WangZYNP2013} with the mode energy $\Omega$ scaling with the superconducting transition temperature $T_c$ in the way $\Omega\approx4.3k_BT_c$. This has been naturally explained as the consequence of the $s^\pm$-pairing gap.

It seems that the model of $s^\pm$-pairing is so far so good for the systems with both electron and hole pockets. However, new challenges come out for some later discovered FeSe-based systems, such as the FeSe monolayer thin film\cite{XueQKCPL}, (Li$_{1-x}$Fe$_x$)OHFeSe\cite{ChenXH,German},
etc., which seem to show the absence of hole pockets in the center of the Brillouin zone\cite{ZhouXJ,FengDL}. The key question is whether we still have sign reversal gaps among the electron pockets. If it exists, what is the configuration of the gap pattern, two candidates would be the nodeless $d$-wave pairing\cite{LeeDH,Chubukov} and the bonding-anti-bonding $s^\pm$ pairing\cite{Mazin2011}. Recently we have adopted a proposal\cite{HirschfeldPRB} for measuring the gap sign and found the evidence of sign-reversal gaps\cite{DuZYNP} in the system (Li$_{1-x}$Fe$_x$)OHFe$_{1-y}$Zn$_y$Se. This method is relying on the determination of a sophisticated quantity associated with the real part of antisymmetrized inter-band Fourier transformed (FT-) QPI. It is expected that this quantity will be coherently enhanced in the region between two gaps. Furthermore a careful calibration is needed to obtain the phase message by implementing the phase-correction method\cite{FeSe}.

Very recently, another method is proposed to judge the gap sign problem in LiFeAs with both electron and hole pockets. Namely the phase information can be validly extracted from the impurity induced bound states\cite{arXiv1,arXiv2}. It seems that this new method is sensitive and straightforward. In present work, we operate QPI measurements around one single impurity in Zn-doped (Li$_{1-x}$Fe$_x$)OHFeSe and also in a large area with multiple impurities in (Li$_{1-x}$Fe$_x$)OHFeSe. Applying this new method for one single impurity, we find out the robust proof of gap sign reversal directly visualized on the two electron pockets. Furthermore, in a system with multiple and identical impurities, we can recover the similar results as the case of one single impurity when carrying out the phase-correction of these impurities. Our results indicate that the unconventional Cooper pairing in (Li$_{1-x}$Fe$_x$)OHFeSe is originated from the on-site Coulomb interaction, as previously proposed in the FeAs-based superconductors.

\section{EXPERIMENTAL METHOD}

The single crystals of (Li$_{1-x}$Fe$_x$)OHFeSe and (Li$_{1-x}$Fe$_x$)OHFe$_{1-y}$Zn$_y$Se are synthesized by hydrothermal ion-exchange method\cite{ChenXH,Dongxiaoli,DuZYNP}. The value of $y$ in (Li$_{1-x}$Fe$_x$)OHFe$_{1-y}$Zn$_y$Se analyzed by X-ray energy dispersive spectrum is about 2\%. The DC magnetization at 20 Oe shows that the critical temperatures of (Li$_{1-x}$Fe$_x$)OHFeSe and Zn-doped samples are 36.4 K and 33.4 K respectively.

STM/STS measurements are carried out by a scanning tunneling microscope (USM-1300, Unisoku Co., Ltd.) with the ultra-high vacuum, low-temperature and high-magnetic field. All the samples were cleaved at room
temperature in an ultra-high vacuum of 1$\times$$10^{-10}$ Torr, and then transferred into the low-temperature microscope head immediately. The electrochemically etched tungsten tips were used during all the measurements. To raise the signal-to-noise ratio in $dI/dV$ spectra, a typical lock-in technique was used with an ac modulation of 0.4 mV at 987.5 Hz. All data in the paper were taken at 1.5 K.

\section{RESULTS}

\subsection{Theoretical model of bound-state based phase-referenced QPI}

The Bogoliubov quasi-particles with the momentum and energy ($\mathbf{k}$,$E$) can be elastically scattered by defects to another state ($\mathbf{k'}$,$E$), forming a standing wave with the wave vector $\mathbf{q} = \mathbf{k}' -\mathbf{k}$. Such standing waves can be easily observed by STM in the real space from QPI measurement. The measured differential conductance mapping $g(\mathbf{r},E)$ is proportional to the local density of state LDOS, neglecting the spatial variations of the tunneling matrix. Then the detailed information in $q$-space can be obtained by applying the Fourier transformation to QPI data, which reflects the scatterings in $k$-space. The obtained FT-QPI $g(\mathbf{q},E)$ is a complex value, and can be expressed as $g(\mathbf{q},E)=|g(\mathbf{q},E)|e^{i\theta_{\mathbf{q},E}}$ with $\theta_{\mathbf{q},E}$ the phase.

Recently a phase-referenced QPI method was proposed to determine the gap symmetry, and the model is based on the phase-referenced QPI near one single impurity as\cite{arXiv1,arXiv2}
\begin{equation}
g_r(\mathbf{q},+E) = |g(\mathbf{q},+E)|,
\end{equation}
\begin{equation}
g_r(\mathbf{q},-E) = |g(\mathbf{q},-E)|\cos(\theta_{\mathbf{q},-E}-\theta_{\mathbf{q},+E})
\end{equation}
As one can see, $\theta_{\mathbf{q},E}$ in the above equations is used as a referenced phase when compared to $\theta_{\mathbf{q},-E}$. Then the phase difference term can be expressed as
\begin{equation}
\begin{split}
&\cos(\theta_{\mathbf{q},-E}-\theta_{\mathbf{q},+E}) \\
&\  = \cos\theta_{\mathbf{q},-E}\cos\theta_{\mathbf{q},+E}+\sin\theta_{\mathbf{q},-E}\sin\theta_{\mathbf{q},+E} \\
&\ =\frac{\mathrm{Re} [g(\mathbf{q},-E)]}{|g(\mathbf{q},-E)|}\frac{\mathrm{Re} [g(\mathbf{q},+E)]}{|g(\mathbf{q},+E)|} \\
&\ \ \ \ \ +\frac{\mathrm{Im} [g(\mathbf{q},-E)]}{|g(\mathbf{q},-E)|}\frac{\mathrm{Im} [g(\mathbf{q},+E)]}{|g(\mathbf{q},+E)|},
\end{split}
\end{equation}
where Re represents the real part of the complex function $g(\mathbf{q},E)$, and Im represents imaginary part. According to the newly suggested treatment method\cite{arXiv1,arXiv2}, $g_r(\mathbf{q},+E)$ is always taken as positive (see Eq.~(1)). Specially, for the nonmagnetic impurity within $s^\pm$ paring symmetry, the integral signal of $g_r(\mathbf{q},-E)$ from sign reversed scattering will change into negative values, with the intensity peak emerging near the impurity induced bound state energy as well.

Furthermore, this method can be also used in a system with multiple impurities\cite{arXiv1}, and here the measured differential conductance mapping $g(\mathbf{r},E)$ can be converted into
\begin{equation}
g(\mathbf{r},E) =\sum_j g{_s}(\mathbf{r} - \mathbf{R}_j,E),
\end{equation}
where $\mathbf{R}_j$ is the location of the $j^\mathrm{th}$ impurity, and then $g{_s}(\mathbf{r}- \mathbf{R}_j,E)$ is the differential conductance mapping by moving the center of the $j^\mathrm{th}$ impurity to the origin. Here the subscript ¡®s¡¯ means for single impurity. After a brief mathematical operation\cite{arXiv1}, we can recover $g{_s}(\mathbf{q},E)$ from $g(\mathbf{q},E)$, namely
\begin{equation}
g{_s}(\mathbf{q},E)=\frac{g(\mathbf{q},E)}{\sum_j e^{-i\mathbf{q}\cdot\mathbf{R}{_j}}}.
\end{equation}
From the equation above, with no need of complicated calculation concerned, the denominator $\sum_j e^{-i\mathbf{q}\cdot\mathbf{R}{_j}}$ is only a complex parameter determined by the location of each impurity and can be calculated from experimental data. As a result, it could be a practical way to obtain $g{_s}(\mathbf{q},E)$ directly from $g(\mathbf{q},E)$, then $g_r(\mathbf{q},\pm E)$ can be calculated from Eqs.~(1) and (2) by replacing the information of $g(\mathbf{q},E)$ by that of $g{_s}(\mathbf{q},E)$. Above all, we can conclude that this phase sensitive method is applicable both for the case of one single impurity and also for a system with multiple impurities, if a careful calibration is done.

\subsection{The bound-state based phase-referenced QPI method applied on one single impurity situation}

\begin{figure}
\includegraphics[width=8cm]{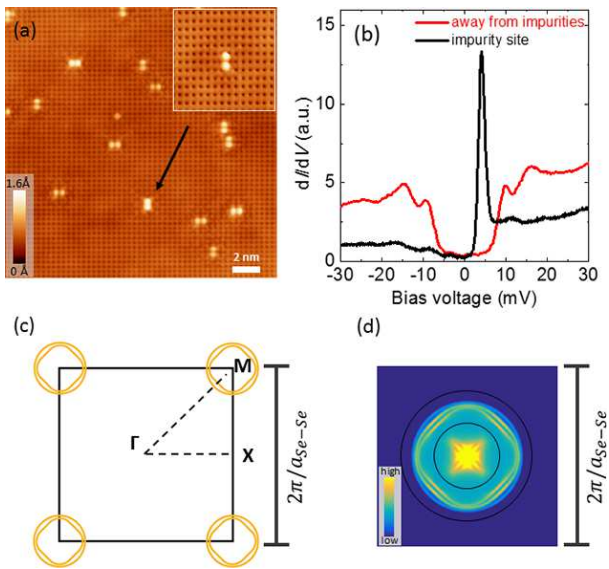}
\caption {(Color online) (a) A typical topographic image of Se terminated layer in (Li$_{1-x}$Fe$_x$)OHFe$_{1-y}$Zn$_y$Se measured with bias voltage $V_{b}=30$ mV and tunneling current $I_t=50$ pA. The arrow indicates a well-isolated Fe-site impurity, and the inset shows the rescanned image with higher resolution around this impurity ($V_{b}=30$ mV, $I_t=100$ pA). (b) Tunneling spectra measured at the center of the impurity marked by the arrow in (a) and far away from impurities. (c) The schematic Fermi surfaces in 2-Fe Brillouin zone. The two electron pockets with moderate hybridization are located around each M point, and the sizes of the electron pockets are determined from the measured FT-QPI results\cite{DuZYNP}. (d) The simulation of FT-QPI intensity by using self-correlation for (c). For clarity, only the central pattern with small-$q$ is presented here. The selected region between the two solid circles is in the region of $0.37\pi/a_{\mathrm{Se}-\mathrm{Se}}<q<0.74\pi/a_{\mathrm{Se}-\mathrm{Se}}$, which is used as the integral region for bound-state based phase-referenced QPI (see text).
}\label{fig1}
\end{figure}

Figure~\ref{fig1}(a) shows a typical atomically resolved topography of a (Li$_{1-x}$Fe$_x$)OHFe$_{1-y}$Zn$_y$Se sample after cleavage, and the Se terminated surface shows a square lattice with lattice constant close to 3.7 ${\AA}$. The single impurity on Fe sites shows a topography with a dumbbell shape\cite{DuZYNP}. The inset of Fig.~\ref{fig1}(a) shows the topographic image with a well-isolated impurity in the center. The spectrum measured at an impurity-free position is shown in Fig.~\ref{fig1}(b), and the spectrum is featured by a standard ``U'' shape indicating an $s$-wave pairing without any nodes crossing the Fermi surfaces. One can also easily distinguish two gaps $\Delta_1 \approx 14$ meV and $\Delta_2 \approx 8.5$ meV from the spectrum, therefore, the material behaves as a multi-gap superconductor like Ba$_{0.6}$K$_{0.4}$Fe$_2$As$_2$ \cite{Shanlei}, LiFeAs\cite{Hanaguri2012}, etc. Moreover, the impurity induced bound state peaks appear at around $ E_B=\pm4$ meV, although the peak has a very weak and almost negligible amplitude at the negative energy. This impurity is proved to be a nonmagnetic one evidenced by the non-shift of the peak energy under the magnetic field of 11 T\cite{DuZYNP}.

As revealed by previous angle resolved photo-emission (ARPES)\cite{ZhouXJ,FengDL} and STM/STS measurements\cite{DuZYNC}, the hole pockets are absent near $\Gamma$ point of the Brillouin zone in (Li$_{1-x}$Fe$_x$)OHFeSe or (Li$_{1-x}$Fe$_x$)OHFe$_{1-y}$Zn$_y$Se\cite{DuZYNP}. The schematic Fermi surfaces are shown in Fig.~\ref{fig1}(c). The intensity is assumed as a constant everywhere around the Fermi surfaces and then we simulate the FT-QPI pattern by using self-correlation for Fig.~\ref{fig1}(c), and present the small-$q$ results in Fig.~\ref{fig1}(d). The two electron pockets can give rise to two sets of the intra-pocket scattering and one set of the inter-pocket scattering. The region between the solid circles in the figure contains the main scattering intensity of the intra- and inter-pockets scattering, which will be used as the integral area in the bound-state based phase-referenced QPI calculations.

\begin{figure}
\includegraphics[width=7cm]{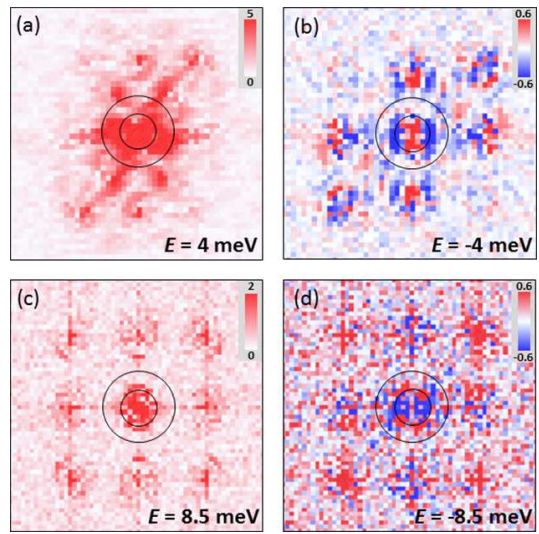}
\caption {(Color online) Bound-state based phase-referenced QPI patterns $g_r(\mathbf{q},E)$ at different energies. These data obtained from the QPI images measured with $64\times64$ pixels in the area with the topography shown in the inset of Fig.~\ref{fig1}(a). The integral process in each figure is carried out in the region between two solid circles which are the same as those shown in Fig.~\ref{fig1}(d).
} \label{fig2}
\end{figure}

The QPI patterns were measured at different energies between $-24$ meV and $+24$ meV in the region whose topography is shown in the inset of Fig.~\ref{fig1}(a), and the nonmagnetic impurity is located in the center of the image. After carrying out some mathematical procedures based on Eqs.~(1) and (2), we can get a series of phase-referenced QPI patterns from the raw data. In Fig.~\ref{fig2}, we present resultant patterns of $g_r(\mathbf{q}, E)$ at typical energies $\pm E=\pm 4$ meV and $\pm 8.5$ meV, which are at the impurity bound state energies and the smaller gap $\pm\Delta_2$, respectively. Obvious twofold symmetry can be observed in the resultant bound-state based phase-referenced  FT-QPI images . The reason is that the Fe-site impurity sits just under the midpoint between the two nearest neighbored Se atoms on the surface, which naturally lowers down the fourfold symmetry of the square lattice and this can get support from the topographic image near the impurity. The two circles in each figure have the same sizes as those in Fig.~\ref{fig1}(d), and the region between them contains the main scattering intensity of the intra- or inter-pockets scattering. It is not strange that $g_r(\mathbf{q}, E)$ at some positive energy is positive everywhere, because it is the absolute value according to Eq.~(1). However, negative values seem to be dominant in the concerned region between two circles at $E=-4$ meV, which indicates that the selected area contains the sign reversal inter-pocket scattering. The area with negative value of $g_r(\mathbf{q}, E<0)$ shrinks when $|E|$ increases, and then the positive and negative areas are almost mixed and balanced near $-8.5$ meV. By the way, we have a feeling that the phase-referenced method based on the bound states\cite{arXiv1,arXiv2} may be only applicable at and near by the bound state peak energy, not effective like the method sensitive to the energies between two gaps\cite{HirschfeldPRB} since there are some differences between the two methods.

\begin{figure}
\includegraphics[width=9cm]{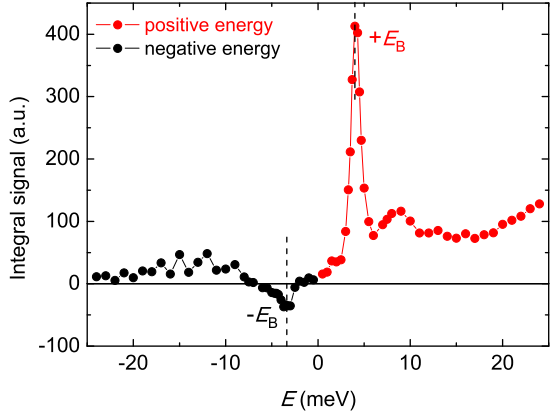}
\caption {(Color online) The calculated integral signal of $g_r(\mathbf{q},E)$ versus energy. The integrated area in $q$-space is restricted between the two circles shown in Fig.~\ref{fig2}.
} \label{fig3}
\end{figure}

Inferred from the theoretical models and previous experimental results\cite{arXiv1,arXiv2}, the absolute value of phase-referenced QPI signal is enhanced significantly when the energy is close to the in-gap bound state. The major difference is the sign of the phase-referenced QPI signals near the bound state peak at the negative energies for different kinds of impurities in superconductors with different gap symmetries, i.e., negative for nonmagnetic impurity in a superconductor with sign reversal gaps and positive for magnetic impurity in a superconductor with sign preserved gaps\cite{arXiv2}. To quantitatively describe the feature of phase-referenced QPI in the sample, we calculate the integrals of $g_r(\mathbf{q},E)$ over the selected area at different energies ranging from $- 24$ meV to $+ 24$ meV, and plot the experimental result in Fig.~\ref{fig3}. The peaks for integral signal near the impurity bound states ``$\pm E_B$'' energies have a clear sign reversal from the positive to negative energy sides, which is consistent with the result from nonmagnetic impurity in $s^\pm$ model\cite{arXiv2}. Accordingly, we believe this is another proof of sign reversal superconducting gaps between the two electron pockets in (Li$_{1-x}$Fe$_x$)OHFe$_{1-y}$Zn$_y$Se, which is totally consistent with our previous conclusion\cite{DuZYNP}.

\subsection{Control experiment on another kind of single impurity}

\begin{figure}
\includegraphics[width=7.5cm]{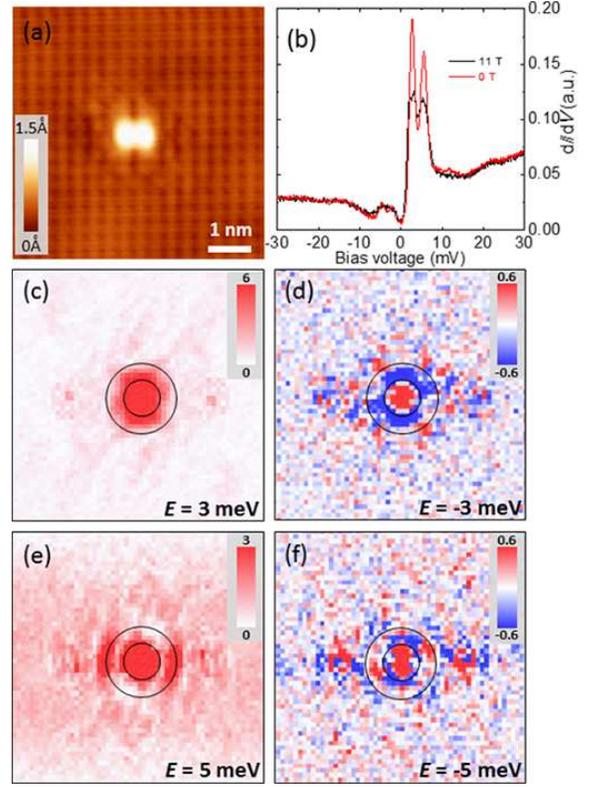}
\caption {(Color online) (a) Topography of another single impurity located at the center of the image with the FOV of 6 nm $\times$ 6 nm ($V_{b}=30$ mV, $I_t=100$ pA). (b) Tunneling spectra measured on the impurity site under the magnetic fields of 0 T and 11 T, respectively. (c)-(f) Phase-referenced QPI patterns at $E=\pm 3$ meV and $\pm 5$ meV. The two circles adhered to each figure are the same sizes as depicted in Fig.~\ref{fig1}(d).
} \label{fig4}
\end{figure}

In order to reinforce the reliability of the analyzing method and also conclusions above, we have carried out a control experiment on another kind of impurity. Figure~\ref{fig4}(a) shows a single impurity which is well located at the center in a field of view (FOV) with dimensions of 6 nm $\times$ 6 nm. The impurity pattern is dumbbell shaped as well. In Fig.~\ref{fig4}(b), we show the spectra measured at the impurity site under magnetic fields of 0 T and 11 T, respectively. At zero field, one can see that two pairs of bound states peaks emerge at $\pm E_{B1}=\pm2.7$ meV and $\pm E_{B2}=\pm5.6$ meV, which is different from the impurity in the previous subsection with only one pair of impurity bound states. The high magnetic field does not shift the peak positions of the in-gap states, manifesting the non-magnetic character of this impurity\cite{DuZYNP}.

\begin{figure}
\includegraphics[width=9cm]{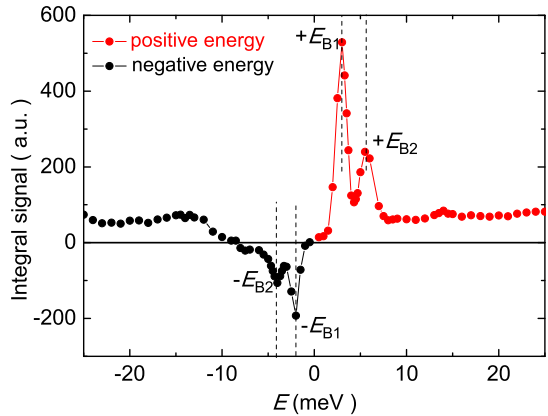}
\caption {(Color online) The integral of $g_r(\mathbf{q},E)$ versus $E$ varying from $- 25$ meV to $+ 25$ meV for the impurity shown in Fig.~\ref{fig4}(a). The two pairs of peaks located at $\pm E_{B1}$, $\pm E_{B2}$ are attributed to the impurity induced bound states.
} \label{fig5}
\end{figure}

Subsequently, a set of differential conductance mappings were measured in the region shown in Fig.~\ref{fig4}(a). The phase-referenced QPI patterns can be calculated from the measured data, and four of them are presented in Figs.~\ref{fig4}(c)-(f). One can clearly see that most of the values between the two circles are negative at the energies of $-3$ meV and $-5$ meV, which are close to the in-gap bound state energies. Then we calculated the integral signals over the area between the two circles, and the energy evolution of the signal is plotted in Fig.~\ref{fig5}. From the resultant curve, these two pairs of the integral signal peaks are located near in-gap state energies marked by ``$\pm E_{B1}$'' and ``$\pm E_{B2}$'', with a sign changing at the positive and negative energy sides. Clearly, the experimental results of the two different impurities are consistent with the theoretical calculation in which there exists a sign reversal gap between the two electron pockets in (Li$_{1-x}$Fe$_x$)OHFe$_{1-y}$Zn$_y$Se.

\subsection{Same method applied on multiple impurities}

\begin{figure}
\includegraphics[width=8cm]{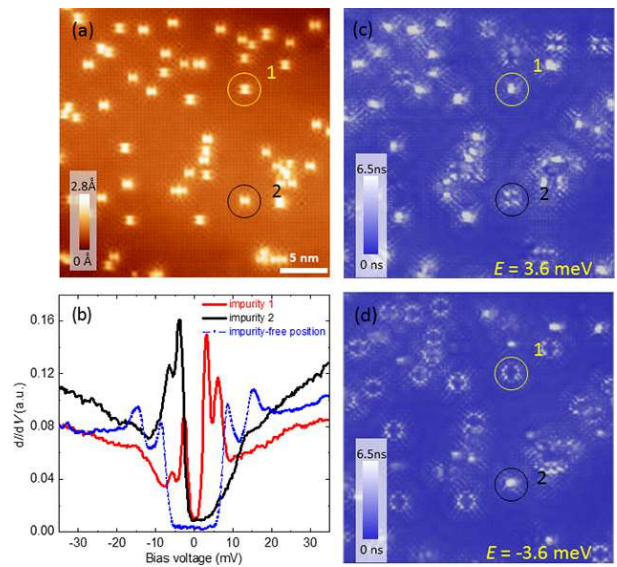}
\caption {(Color online) (a) Topography of a 28 nm $\times$ 28 nm area with plenty of dumbbell shaped impurities in (Li$_{1-x}$Fe$_x$)OHFeSe sample ($V_{b}=40$ mV, $I_t=100$ pA). (b) Spectra measured at the centers of two different impurities [marked as 1 and 2 in (a)] and at an impurity-free position.  (c),(d) Differential conductance mappings measured at $\pm3.6$ mV in the same area as the topography shown in (a). As one can see, impurities in (a) can be mainly categorized into two kinds.
} \label{fig6}
\end{figure}

In the next, we present the data of a new round of experiments on a Zn-free sample with multiple impurities. As shown in Fig.~\ref{fig6}(a), plenty of impurities are witnessed in a FOV of 28 nm $\times$ 28 nm and all defects show similar dumbbell-shapes. Tunneling spectra measured at the centers of the two different impurities and at an impurity-free position are presented in Fig.~\ref{fig6}(b). The tunneling spectrum measured at the impurity-free position is featured by a ``U'' shape, which indicates a nodeless gap feature. The clear and sharp coherence peaks reveal the double gaps in the Zn-free samples, and the feature is also very close to the one measured on the Zn-doped samples. Impurities in FOV can be mainly categorized into two kinds, and the symbols for these two kinds are impurity 1 (marked by yellow circle) and impurity 2 (marked by black circle), respectively. As for impurity 1, the impurity-induced bound state peaks at the positive energies are much stronger than those at the negative energies, and the situation is reversed with respect to impurity 2. In Figs.~\ref{fig6}(c,d), we display differential conductance mappings measured at energies close to the in-gap state with the smaller peak energy. One can clearly see that there exists an obvious difference between the QPI patterns induced by impurity 1 and impurity 2. Arising from the synthesis procedure of hydrothermal ion-exchange method with many kinds of elements in (Li$_{1-x}$Fe$_x$)OHFeSe, these two different kinds of impurities may come from the vacancies of Fe or the substitution of Fe-sites by atoms of other elements, probably the Li atoms.

\begin{figure}
\includegraphics[width=8cm]{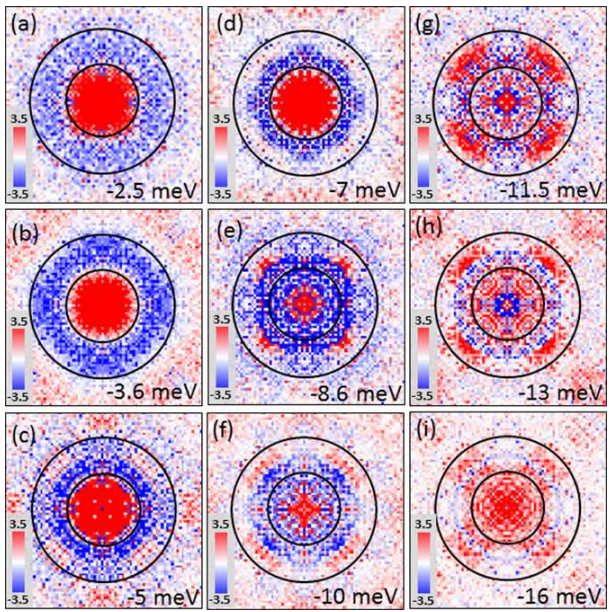}
\caption {(Color online) The phase-referenced QPI patterns of $g_s(\mathbf{q},E)$ at negative energies for impurity 1. The side length and the selected region between the two circles for integration are the same as those shown in Fig.~\ref{fig1}(d).
} \label{fig7}
\end{figure}

\begin{figure}
\includegraphics[width=9cm]{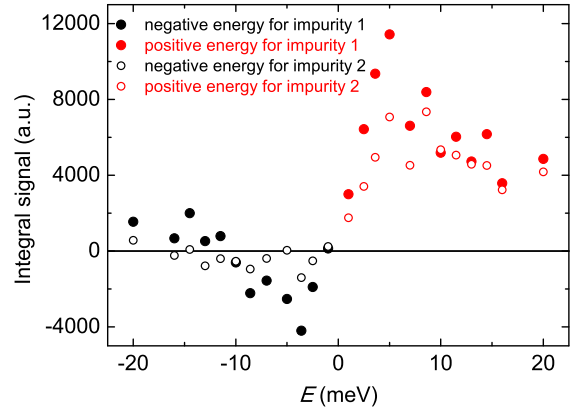}
\caption {(Color online) The integral signal of phase-referenced QPI for impurity 1 and impurity 2. The extrema emerge at the energies close to the impurity bound states.
} \label{fig8}
\end{figure}

Therefore, if we want to recover $g{_s}(\mathbf{q},E)$ from the QPI measurements for the large area with multiple impurities, it is necessary to mask out one kind of impurities with another kind left. Thus we need to deduct the contribution of that kind of impurities. For that purpose, the values in the circle surrounding one kind of impurity (1 or 2) with a radius 1.6 nm are substituted by the average value of the whole differential conductance mapping, as a result there exist only one kind of identical impurities of interests in the masked mapping. Then we can get a series of phase-referenced QPI patterns referring to Eq.~(5) and then Eqs.~(1) and (2). In order to figure out the energy evolution of $g_r(\mathbf{q},E)$, we present a series of patterns at the negative energies varying from $- 2.5$ meV to $- 16$ meV for impurity 1 as shown in Fig.~\ref{fig6}(a). The patterns at positive energies are not presented here because they are nothing but the absolute values of FT-QPI without extra phase-related information. As we know, the FT-QPI results have some diffuse weight arising from the long-range disorders in real space, so that the pattern with very small $q$ which is concentrated within the inner circle could be complex and difficult to analyse. As mentioned above, the selected region between two circles will cover the main scattering intensity of the intra- or inter-pockets scattering. One can clearly see that most of the values in the selected area are negative when the energies are close to impurity-induced in-gap state energies of impurity 1 [Fig.~\ref{fig7}(b-d)]. In Fig.~\ref{fig7}(f), it is obvious that there are two neighbored contours with positive and negative values respectively in the selected region, which may be from the different kinds of scatterings if the gap changes its sign for the two electron pockets. When the energy exceeds the larger gap, the positive signals begin to dominate, which can be easily understood as due to  the signal from the normal state. We then plot the energy dependent integral signals of $g_r(\mathbf{q},E)$ for these two kinds of impurities in Fig.~\ref{fig8}. As one can see, the signal reaches its extrema at the energies close to the bound state peaks, meanwhile, it does have a sign change for positive and negative energies. This is consistent with the theoretical prediction for the nonmagnetic impurities in an $s^\pm$ pairing superconductor\cite{arXiv2}. Thus we have successfully recovered the $g{_s}(\mathbf{q},E)$ in a system with multiple impurities, giving strong support for the sign reversal gaps between the two electron pockets in (Li$_{1-x}$Fe$_x$)OHFeSe.

\section{DISCUSSION}

Compared to the case of one single impurity, QPI measurements for a large area with multiple impurities can give us a high resolution in $\mathbf{q}$-space and we can obtain more details of the QPI scatterings from the Fermi surfaces. Between the two circles in Fig.~\ref{fig1}(d), there are three scattering channels, and two of which are sign-reserved and one is sign-changed. Roughly speaking, these three scattering channels will mix together and then it may hinder us from identifying the sign-changed scattering. In fact, specially for the bound state peak at $-3.6$ meV, the selected region is almost covered by the negative sign, thus indicating the existence of sign-reversal gaps. As we understand, the scattering of the Bogoliubov quasi-particles with energy $E_k$ and wavevector $k$ in a superconductor can be characterized by the coherence factors\cite{Hoffman}, namely
\begin{equation}
\begin{split}
u_k &= \frac{\Delta_k}{|\Delta_k|}\sqrt{\frac{1}{2}\left(1+\frac{\varepsilon_k}{E_k}\right)},\\
v_k &= \sqrt{1-|u_k|^2},
\end{split}
\end{equation}
where $|u_k|^2$ and $|v_k|^2$ are the probabilities that Cooper pairs unoccupy and occupy the $\pm k$ state, and $\varepsilon_k$ is the kinetic energy. Within the Fermi's golden rule, the scattering probability from $k$ to $k'$ is roughly proportional to $C(k,k') = |u_ku_{k'}-v_kv_{k'}|^2$ for the scalar potential\cite{2009PRB,Hoffman}. Provided that the scatters are non-magnetic, the value of $C(k,k')$ for the sign-changed scattering will be much larger than the one for the sign-reserved scattering at the low excited energy within superconducting gap. Therefore, we can get the strong signal mainly from the sign-changed inter-pockets scattering channel, the other two scattering channels within the sign-reserved gap should be very weak.

From the experimental data of phase-referenced QPI, we can find that the integral of $g{_r}(\mathbf{q},E)$ has a sign changing of the signal peak between the positive and negative energies near the impurity bound states. We can also notice that the signals at the high energies become positive disregard in the positive or negative energy sides, as shown in Figs.~\ref{fig3}, \ref{fig5} and \ref{fig8}. It should be noted that the sign reversal of phase-referenced QPI signal is based on the phase change originated from the scattering of Bogoliubov quasi-particles within the superconducting gap. However, the situation for the normal state should be different, i.e., the phase angles of $\theta_{\mathbf{q},-E}$ and $\theta_{\mathbf{q},E}$ should be similar for normal state quasi-particles. In another words, at high energies beyond superconducting gaps, $C(k,k')$ will tend to be a constant 1 for scattering of both sign-reversed and reserved processes. Hence, it is not strange that the integral signals of $g{_s}(\mathbf{q},\pm E)$ become a positive value when $|E|$ is much larger than gap values, as in the normal state.

In our recent work\cite{DuZYNP}, we have obtained an elusive quantity, the real part of antisymmetrized FT-QPI, which is defined as $\delta g^{-} (E) = \sum_{\mathbf{q}}\mathrm{Re}[g(\mathbf{q},+E) - g(\mathbf{q},-E)]$, and is coherently enhanced within the energy region between two gaps\cite{HirschfeldPRB}. It provides us a robust evidence of the sign reversal gaps on the two electron pockets in (Li$_{1-x}$Fe$_x$)OHFe$_{1-y}$Zn$_y$Se\cite{DuZYNP}. This phase-sensitive method is designed for the case of one isolated impurity, as a result that the phase message can not be easily affected by other neighbored impurities in the investigated FOV. Back to the recently proposed approach\cite{arXiv1,arXiv2} used in this paper, namely the bound-state based phase-referenced QPI for one single impurity, it is also very helpful to judge the sign problem of the order parameters near the energy of the impurity state. Furthermore, this new approach is applicable for the system with multiple impurities as well and we have successfully recovered the similar result as the measurements of one single impurity. Both methods are quite useful and they can play as a mutual double check. Our results indicate that the phase referenced QPI can provide an easy and feasible way to detect the gap function of unconventional superconductors.

\section{SUMMARY}

We performed QPI measurements around a single impurity in (Li$_{1-x}$Fe$_x$)OHFe$_{1-y}$Zn$_y$Se at a series of energies. Adopting the newly proposed method of bound-state based phase-referenced QPI, we demonstrate that there exists a sign reversal gap between the two electron pockets, namely the inner and outer circles after folding or hybridization. Furthermore, for the situation in (Li$_{1-x}$Fe$_x$)OHFeSe with multiple impurities, the similar results are also obtained, which proves the validity of the method and conclude again the sign reversal gaps in the system. Considering a practical case, sometimes it may not be easy to find out one well-isolated impurity, so that the phase-referenced QPI measurements applied for multiple impurities seem to be more realistic and thus provide a practical way to detect the gap function of unconventional superconductors. Our results suggest that the FeSe-based superconductors without the hole pockets have a sign reversal of gaps between the two electron pockets, being consistent with the picture of unconventional Cooper pairing mediated by exchanging antiferromagnetic spin fluctuations.

\begin{acknowledgments}

This work was supported by National Key R \& D Program of China (Grant No. 2016YFA0300401), National Natural Science Foundation of China (Grant No. 11534005), and Natural Science Foundation of Jiangsu (Grant No. BK20140015).

\end{acknowledgments}

$^*$ huanyang@nju.edu.cn

$^{\dag}$ hhwen@nju.edu.cn

\end{document}